\documentclass[useAMS,usenatbib]{mn2cls/mn2e}
\usepackage{epsfig}
\def\gtsima{$\; \buildrel > \over \sim \;$}
\def\ltsima{$\; \buildrel < \over \sim \;$}
\def\gsim{\lower.5ex\hbox{\gtsima}}
\def\lsim{\lower.5ex\hbox{\ltsima}}

\def\Msun{{M_\odot}}
\def\Zsun{{Z_\odot}}

\def\lisix{{$^6$Li}}
\def\liseven{{$^7$Li}}

\def\be{\begin{equation}}
\def\ee{\end{equation}}

\def\de{{\mathrm{d}}}
\def\II{_{\rm II}}
\def\gg{_{\gamma \! \gamma}}
\def\nap{N_{\alpha,\rm H}}
\def\nli{N_{\rm 6Li,H}}
\def\emin{E_{\rm min}}
\def\emax{E_{\rm max}}

\title[The $^6$Li plateau]{The puzzling origin of the $^6$Li plateau}
\author[Evoli, Salvadori \& Ferrara]
{Carmelo Evoli\thanks{E-mail: evoli@sissa.it}, Stefania Salvadori  \& Andrea Ferrara \\
SISSA/International School for Advanced Studies, Via Beirut 4, 34100 Trieste, Italy}
\date{}
\pagerange{\pageref{firstpage}--\pageref{lastpage}} 
\pubyear{2008}

\begin{document}

\maketitle
\label{firstpage}

\small

\begin{abstract}
We discuss the \lisix \ abundance evolution within a hierarchical model of Galaxy formation 
which correctly reproduces the [Fe/H] distribution of metal-poor halo stars. Contrary to 
previous findings, we find that neither the level ($^6$Li/H$=6\times 10^{-12}$) 
nor the flatness of the \lisix \ distribution with [Fe/H] can be reproduced 
under the most favourable conditions by any model in which \lisix \ production is tied to a 
(data-constrained) Galactic star formation rate via cosmic ray spallation. Thus, the 
origin of the plateau might be due to some other early mechanism unrelated to star formation.
\end{abstract}

\begin{keywords}
nuclear reactions, nucleosynthesis, abundances - stars: abundances - stars: formation - cosmic rays - galaxies: evolution - cosmology: theory.
\end{keywords}

\section[]{Introduction}

The relative abundance of light elements synthesized during the big bang nucleosynthesis (BBN)
is a function of a single parameter, $\eta$, namely the baryon-to-photon ratio. Given the WMAP 
constraint $\eta = (6.8 \pm 0.21) \times 10^{-10}$, the light nuclei abundances can be precisely 
predicted by BBN \citep{Spergel07, PDG}. Despite a general agreement with the observed abundances of 
light elements, discrepancies arise concerning Li abundance.  Observationally, the primordial 
abundance of lithium isotopes (\liseven \ and \lisix ), is measured in the atmospheres of Galactic
metal-poor halo stars (MPHS).

Since the first detection by \cite{Spite82}, later confirmed by subsequent works 
\citep{Spite84, Ryan99, Asplund06, Bonifacio07} a \liseven/H $= (1-2) \times 10^{-10}$ 
abundance was deduced, independent of stellar [Fe/H]. The presence of such a \liseven \ plateau 
supports the idea that \liseven \ is a primary element, synthesized by BBN. The measured value, 
however, results of a factor $2-4$ lower than the expected from the BBN
\liseven/H~$=4.27^{+1.02}_{-0.83} \times 10^{-10}$ \citep{Cyburt04}, 
\liseven/H~$=4.9^{+1.4}_{-1.2} \times 10^{-10}$ \citep{Cuoco04}, or 
\liseven/H~$=4.15^{+0.49}_{-0.45} \times 10^{-10}$ \citep{Coc04}.
Recently, \cite{Pinsonneault02} and \cite{Korn06} found that mixing and diffusion processes during stellar 
evolution could reduce the \liseven \ abundance in stellar atmospheres by about 0.2~dex, thus 
partially releasing the tension.
 
A more serious problem arose with \lisix, for which the BBN predicts a value of (\lisix/H)$_{\rm BBN} 
\sim 10^{-14}$. Owing to the small difference in mass between \lisix \ and \liseven, lines from these two 
isotopes blend easily. The detection of \lisix \ then results quite difficult since the predominance of 
\liseven. Recently, high-resolution spectroscopic observations measured the \lisix \ abundance in 24 MPHS
\citep{Asplund06}, revealing the presence of a plateau \lisix/H$=6\times 10^{-12}$ for 
$-3\lsim $[Fe/H]$\lsim -1$. A primordial origin of \lisix \ seems favoured by the presence of the 
plateau; however, the high \lisix \ value observed cannot be reconciled with this hypothesis.

The solutions invoked to overcome the problem were: 
(i) a modification of BBN models \citep{Kawasaki05,Jedamzik06,Pospelov06,Cumberbatch07,Kusakabe07},
(ii) the fusion of $^3$He accelerated by stellar flares with the atmospheric helium \citep{Tatischeff07},
(iii) a mechanism allowing for later production of \lisix \ during Galaxy formation.
The latter scenario involves the generation of cosmic rays (CRs). \lisix, in fact, can be 
synthesized by fusion reactions ($\alpha + \alpha \rightarrow \rm \, ^6Li$) when high-energy CR                
particles collide with the ambient gas. Energetic CRs can either be accelerated by shock 
waves produced during cosmological structure formation processes \citep{Miniati00, Suzuki02, Keshet03} or, 
by strong supernova (SN) shocks along the build-up of the Galaxy. 
In their recent work \cite{Rollinde06} used the supernova rate (SNR) by \cite{Daigne06} to compute the 
production of \lisix \ in the intergalactic medium (IGM).  Assuming that all MPHS form at $z\sim 3$, and 
from a gas with the same IGM composition, they obtained the observed \lisix \ value. Despite the apparent 
success of the model, these assumptions are very idealized and require a closer inspection. 
We revisit the problem using a more realistic and data-constrained approach, based
on the recent model by \cite{Salvadori07} (SSF07), which follows the hierarchical build-up 
of the Galaxy and reproduces the metallicity distribution of MPHS.

\section{Building the Milky Way}

The code GAlaxy Merger Tree \& Evolution (\textsc{gamete}) described in SSF07 
(updated version in \citealt{Salvadori08}) follows the star formation 
(SF)/chemical history of the MW along its merger tree, finally matching 
all its observed properties. 

The code reconstructs the hierarchical merger history of the MW using a 
Monte Carlo algorithm based on the extended Press \& Schechter theory 
 \citep{Press74} and adopting a binary scheme with accretion mass 
\citep{Cole00, Volonteri03}. Looking back in time at any time-step 
a halo can either lose part of its mass (corresponding to a cumulative 
fragmentation into haloes below the resolution limit $M_{\rm res}$) or 
lose mass and fragment into two progenitors. The mass below $M_{\rm res}$ 
accounts for the {\it Galactic Medium} (GM) which represents the mass 
reservoir into which haloes are embedded.
During the evolution, progenitor haloes accrete gas from the GM and 
virialize out of it. We assume that feedback suppresses SF in mini-haloes 
and that only Ly$\alpha$ cooling haloes ($T_{\rm vir}>10^4$~K) contribute 
stars and metals to the Galaxy. This motivates the choice of a 
resolution mass $M_{\rm res}= M_4(z)/10=M(T_{\rm vir}=10^4$~K~$,z)/10$ 
where $M_4(z)$ is the mass corresponding to a virial temperature $T_{\rm vir}=10^4$K 
at redshift $z$. At the highest redshift of the simulation, $z\approx 20$, the gas 
present in virialized haloes, as in the GM, is assumed to be of primordial 
composition. The SF rate (SFR) is taken to be proportional to the mass of gas. 
Following the critical metallicity scenario \citep{Bromm01, Schneider02, Schneider03, Omukai05, Schneider06}
we assume that low-mass (Pop~II/I) SF occurs when the metallicity $Z_{\rm cr}>10^{-5\pm 1}\Zsun$
according to a Larson initial mass function with a characteristic mass $m_\star=0.35 M_\odot$.
At lower $Z$ massive Pop~III stars form with a characteristic mass $m_{\rm PopIII}=200\Msun$, 
i.e. within the pair-instability supernova (SN$_{\gamma\gamma}$) mass range of 
$140-260\Msun$ \citep{Heger02}.  The chemical evolution of both gas in proto-Galactic haloes (ISM) 
and in the GM, is computed by according to a mechanical feedback prescription (see \cite{Salvadori08} for details). 
Produced metals are instantaneously and homogeneously mixed with the gas. 

\begin{figure}
  \centerline{\psfig{figure=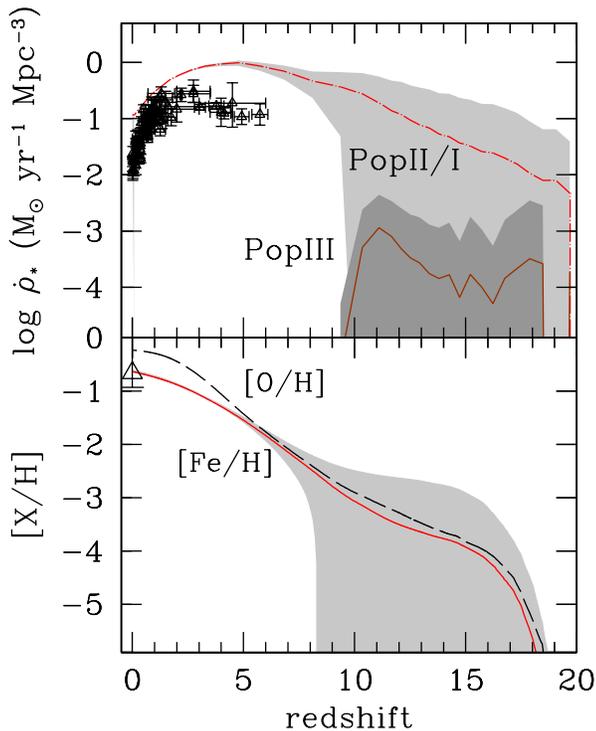,width=8.0cm,angle=0}}
  \caption{{\it Upper panel}: Comoving SFR density evolution for Pop~III (solid line) and Pop~II/I
    stars (dashed line). The curves are obtained after averaging over 100 
    realizations of the merger tree; shaded areas denote       
    $\pm 1\sigma$ dispersion regions around the mean. Points 
    represent the low-redshift measurements of the cosmic SFR by \citet{Hopkins04}. 
    {\it Lower panel}: Corresponding GM iron (solid line) and oxygen 
    (dashed line) abundance evolution.  The point is the measured 
    [O/H] abundance in high-velocity clouds by \citet{Ganguly05}.}
  \label{fig:1}
\end{figure} 

The model free parameters are fixed to match the global properties of the MW and the Metallicity 
Distribution Function (MDF) of MPHS derived form the Hamburg-ESO Survey (Beers \& Christlieb, private communication).
In Fig.~\ref{fig:1} (upper panel) the derived Galactic (comoving) SFR density is shown 
for Pop~III and Pop~II/I stars. Pop~II/I stars dominate the SFR at any 
redshift. Following a burst of Pop~III stars, in fact, the metallicity of the host halo raises to 
$Z>Z_{cr}$: chemical feedback suppresses Pop~III formation in self-enriched 
progenitors. Later on Pop~III stars can only form in those haloes which virialize from 
the GM and so, when $Z_\mathrm{GM}\gsim Z_{cr}$, their formation is totally quenched. 
The above results are in agreement with recent hydrodynamic simulations implementing
chemical feedback  effects (\citealt{Tornatore07}).  The earlier Pop~III disappearance of 
our model ($z \sim 10$) with respect to this study ($z \sim 4$) is a consequence of the biased volume we consider
i.e. the MW environment. As the higher mean density 
accelerates SF/metal enrichment, PopIII stars disappear at earlier times; 
the SFR maximum value and shape, however, match closely the simulated ones. 

In Fig.~\ref{fig:1} (lower panel) we show the corresponding evolution of the GM iron and oxygen 
abundance. As SSF07 have shown that the majority of present-day iron-poor stars ([Fe/H]$<-2.5$) 
formed in haloes accreting GM gas which was Fe-enhanced by previous SN explosions, 
the initial [Fe/H] abundance within a halo is set by the corresponding GM 
Fe-abundance at the virialization redshift.

\section{Lithium production}

To describe the production of \lisix \ for a continuous source of CRs we generalize the classical work 
of \cite{Montmerle77}, who developed a formalism to follow the propagation of an homogeneous 
CR population in an expanding universe, assuming that CRs have been instantaneously produced at some redshift. 

Since the \emph{primary} CRs are assumed to be produced by SNe, the physical source 
function $Q(E,z)$ is described by a power law in momentum: 
\be \label{source}
Q(E,z) = C(z) \frac{\phi(E)}{\beta(E)} \ \mathrm{(GeV/n)}^{-1} \, \mathrm{cm}^{-3} \, \mathrm{s}^{-1}
\ee
with $\beta = v/c$ and
\be
\phi(E) = \frac{E+E_0}{[E(E+2E_0)]^{(\gamma+1)/2}} \ \mathrm{(GeV/n)}^{-1} \, \mathrm{cm}^{-2} \, \mathrm{s}^{-1}
\ee
where $\gamma$ is the injection spectral index and $E_0=939$~MeV and $E$ are, respectively, the rest-mass energy and the 
kinetic energy per nucleon. The functional form of the injection spectrum $\phi(E)$ is inferred from the theory of 
collisionless shock acceleration \citep{Blandford87} and the $\gamma$ value is the one typically associated to the 
case of strong shock. We note however that the results are only very weakly dependent on the spectral slope. 
Finally, $C(z)$ is a redshift-dependent normalization; its value is fixed at each redshift by normalizing $Q(E,z)$ 
to the total kinetic energy transferred to CRs by SN explosions:
\be \label{energetic}
\mathcal{E}_{\mathrm{SN}} (z) = \int_{E_{\mathrm{min}}}^{E_{\mathrm{max}}} E Q(E,z) \de E
\ee
with
\be
\mathcal{E}_{\mathrm{SN}} (z) = \epsilon (1+z)^3 [E_{\II} \mathrm{SNR}_{\II}(z) + E_{\gamma \gamma} \mathrm{SNR}_{\gamma \gamma}(z)]
\ee
where $E\II = 1.2 \times 10^{51}$~erg and $E\gg = 2.7 \times 10^{52}$~erg are, respectively, the average explosion 
energies  for a Type II SN (SN$\II$) and a SN$\gg$; $\epsilon=0.15$ is the fraction of the total energy not emitted in neutrinos
transferred to CRs by a single SN, assumed to be the same for the two stellar populations; SNR$\II$ (SNR$\gg$) is 
the SN$\II$ (SN$\gg$) explosion comoving rate, simply proportional to the Pop~II/I (Pop~III) SFR. 
The efficiency parameter is inferred by shock acceleration theory and confirmed by recent observations of 
SN remnants in our Galaxy \citep{Tatischeff08}.

We now need to specify the energy limits $\emin$, $\emax$ of the CR spectrum produced by SN shock waves 
(eq.~\ref{energetic}). We fix $\emax = 10^6$~GeV, following the theoretical estimate by \cite{Lagage83}. Due to the 
rapid decrease of $\phi(E)$ the choice of $\emax$ does not affect the result of the integration and hence the derived 
$C(z)$ value. On the contrary $C(z)$ strongly depends on the choice of $\emin$: the higher $\emin$, the higher is $C(z)$. 
Since observations cannot set tight constraints on $\emin$, due to solar magnetosphere modulation of low-E CRs, 
we consider it as a free parameter of the model.

Once the spectral shape of $Q(E,z)$ is fixed, we should in principle take in account the subsequent propagation of CRs 
both in the ISM and GM. Following \cite{Rollinde06}, we make the hypothesis that primary CRs escape 
from parent galaxies on a timescale short enough to be considered as immediately injected in the GM without 
energy losses. At high redshift in fact: (i) structures are smaller and less dense (Zhao et al. 2003) implying higher 
diffusion efficiencies (Jubelgas et al. 2006); (ii) the magnetic field is weaker and so it can hardly confine CRs into 
structures. Note also that, besides diffusive propagation of CRs, superbubbles and/or galactic winds could directly 
eject CRs into the GM. 

Under this hypothesis the density evolution of primary CRs only depends on energy losses suffered in the GM. 
The nuclei lose energy mainly via two processes, ionization and Hubble expansion, and they are destroyed by 
inelastic scattering off GM targets (mainly protons). 

We can follow the evolution of $\alpha$-particles (primary CRs) through the transport equation \citep{Montmerle77}
\be \label{alpha}
\frac{\partial \nap}{\partial t} + \frac{\partial}{\partial E} (b \nap)  + \frac{\nap}{T_D} = K_{\alpha p} Q_{\rm ,H}(E,z)
\ee
where $N_{i,H}$ is the ratio between the (physical) number density of species $i$ and  GM protons, 
$n_H(z) = n_{\rm H,0} (1+z)^3$; $Q_{\rm,H} (E,z) \equiv Q(E,z)/n_{\rm H}(z)$ is the normalized physical source function, 
$b \equiv \left( \partial E / \partial t \right)$ is the total energy loss rate adopted from \cite{Rollinde06}, 
$T_D$ is the destruction term as in the analytic fit by \cite{Heinbach95}; finally, 
$K_{\alpha p} = 0.08$ is the cosmological abundance by number of $\alpha$-particles with respect to protons.

We consider \lisix \ as entirely secondary, i.e. purely produced by fusion of GM He-nuclei by 
primary $\alpha$-particles. The physical source function for \lisix \ is given by:
\be \label{qli}
Q_{\rm 6Li}(E,z) = \int \sigma_{\alpha \alpha \rightarrow \rm 6Li}(E,E') n_{\rm He}(z) \Phi_{\alpha}(E',z) \de E'
\ee
where $E'$ and $E$ are respectively the kinetic energies per nucleon of the incident particle and of the produced 
\lisix \ nuclei, and $\Phi_{\alpha}(E',z) = \beta(E') N_\alpha (E',z)$ the incident $\alpha$-particle flux. 
Making the approximation $\sigma_{\alpha \alpha \rightarrow \rm 6Li}(E,E') = \sigma_l (E) \delta (E-E'/4)$ 
\citep{Meneguzzi71} and defining $Q_{\rm 6Li,H} \equiv Q_{\rm 6Li}/n_{\rm H}$, the eq. (\ref{qli}) becomes
\be
Q_{\rm 6Li,H} (E,z) = \sigma_l(E) K_{\alpha p} n_{\rm H}(z) \Phi_{\alpha \rm ,H}(4E,z)
\ee
where the cross section $\sigma_l(E)$ is given by the analytic fit of \cite{Mercer01}:
\be
\sigma_l (E) \sim 66 \exp \left( -\frac{E}{4 \ \rm MeV} \right) \ {\rm mb}
\ee

We can now write a very simple equation describing the evolution of \lisix:
\be \label{lithium}
\frac{\partial \nli}{\partial t} = Q_{\rm 6Li,H}(E,z)
\ee
in this case, in fact, destruction and energy losses are negligible since their time scales are very long 
with respect to the production time scale \citep{Rollinde05}.

The solution of the coupled eqs. (\ref{alpha})-(\ref{lithium}) gives \lisix/H at any given redshift $z$.

\section[]{Results}

The system of equations introduced in the previous Sec.~are solved numerically using a Crank-Nicholson
implicit numerical scheme \citep{NR}. Because of its stability and robustness implicit schemes are used 
to solve transport equations in most CRs diffusion problems \citep{Strong98}.

We test the accuracy of our code by studying a simplified case in which an analytic solution can be
derived and compared with numerical results. To this aim we assume that: (i) both energy losses and destruction 
of primary CRs in the GM can be neglected; (ii) the physical energy density injected by SNe 
is constant, ${\cal E}_{\rm SN} \sim 7.4 \times 10^{-27}$~GeV~cm$^{-3}$~s$^{-1}$, in the 
redshift range $z>3$. 
It is worth noting that the above hypothesis conspire to give an upper limit to the exact  
solution, thus providing an estimate of the maximum achievable \lisix \ abundance. Under these    
approximations, the source spectrum defined in eq.~(\ref{source}) becomes:
\be
Q(E,z) = \rm 6.4 \cdot 10^{-29} \frac{\phi(E)}{\beta(E)} \ \mathrm{(GeV/n)}^{-1} \, \mathrm{cm}^{-3} \, \mathrm{s}^{-1}
\ee
and eqs. (\ref{alpha})-(\ref{lithium}) can be solved. We find
\be
\nap (z)= 39.6 \, (1+z)^{-9/2} 
\ee
and 
\be\label{lithium_analytically}
\nli (z)= 8.2 \times 10^{-11} (1+z)^{-3} 
\ee

\begin{figure}
    \centerline{\epsfig{figure=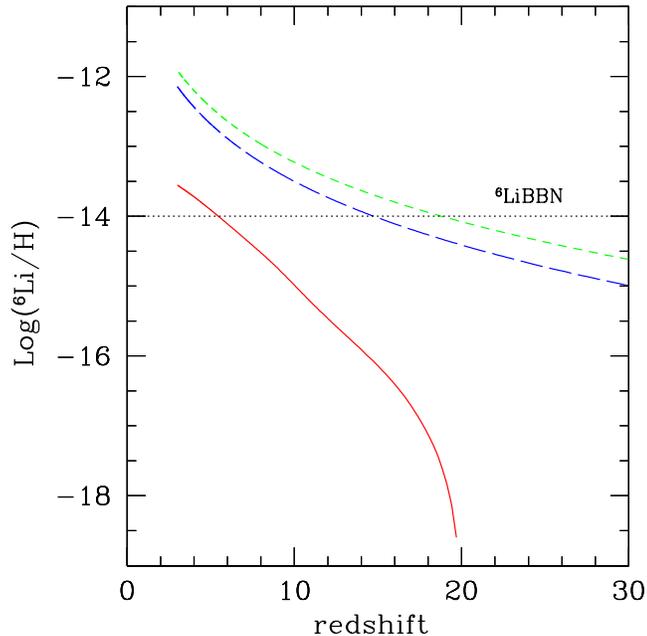,width=9.0cm,angle=0}}
    \caption{Redshift evolution of GM \lisix/H abundance for the analytical (green short-dashed line) 
    and numerical solution (overlapped) of a simplified model with no energy losses and destruction, 
    and ${\cal E}_{\rm SN} = 7.4 \times 10^{-27}$~GeV~cm$^{-3}$~s$^{-1}$ for $z>3$, the same model 
    including energy loss/destruction (blue long dashed line), the fiducial model with realistic SNR, 
    $\epsilon = 0.15$ and $\emin=10^{-5}$~MeV (red solid line).}
\label{fig:2}
\end{figure}

From Fig.~\ref{fig:2} we conclude that the analytical solution for the GM \lisix \ abundance 
(eq.~\ref{lithium_analytically}) is perfectly matched by the numerical\footnote{This solution 
represents an upper limit for the \cite{Rollinde06} model, as inferred from 
their Fig.~2.} one.
Also shown are the numerical solutions obtained by relaxing first the hypothesis (i) and then (i) + (ii).
Not unexpectedly, the inclusion of energy losses and destruction term into eq.~(\ref{alpha})  
affects only slightly the result, as the typical time-scales of such processes are longer than the 
\lisix \ production one. 

A realistic injection energy, on the contrary, has a strong impact on the predicted  
shape and amplitude of the \lisix \ evolution. In fact, the SFR, and consequently ${\cal E}_{\rm SN}$,
is an increasing function of time in the analyzed redshift range (Fig.~1 upper panel). The maximum
${\cal E}_{\rm SN}$ we can obtain by using the SNR derived from the curve in Fig.~1, a realistic energy transfer 
efficiency $\epsilon = 0.15$, and $\emin=10^{-5}$~GeV \citep{Rollinde06}, is
${\cal E}_{\rm SN}^{max}\sim 8.6 \times 10^{-28} < 7.4 \times 10^{-27}$~GeV~cm$^{-3}$~s$^{-1}$.
Note that the \lisix/H abundance at $z=3$ results more than 1 order of magnitude smaller than the value of 
the simplified case. In the following, we will refer to this physical model as our fiducial model. 

\begin{figure}
    \centerline{\epsfig{figure=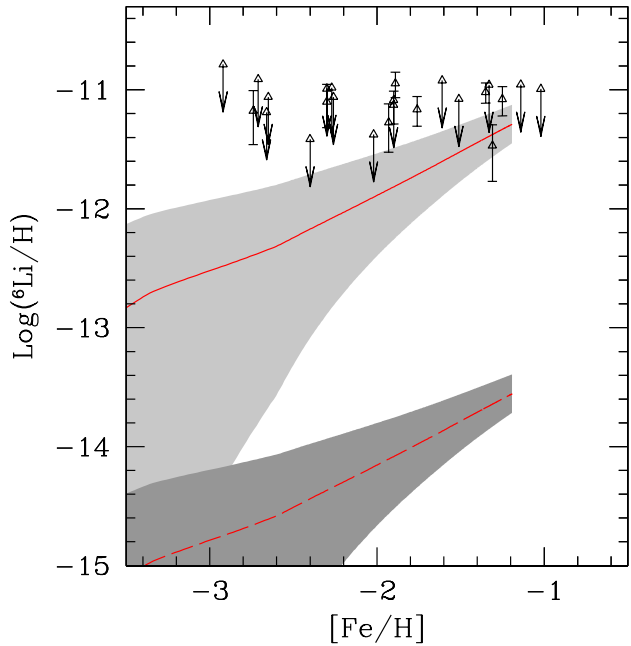,width=9.0cm,angle=0}}
    \caption{Redshift evolution of \lisix/H vs [Fe/H] for the fiducial model ($\epsilon = 0.15$, 
      $\emin=10^{-5}$~GeV/n, dashed line) and for the maximal model ($\epsilon = 1$,   
      $\emin = 10$~MeV/n, solid line). Shaded areas denote $\pm 1\sigma$ dispersion regions around the mean. } 
\label{fig:3}
\end{figure}

We now use the [Fe/H] predicted by GAMETE (Fig.~1, lower panel) to convert redshift into [Fe/H] values and 
derive the GM \lisix \ vs [Fe/H]. According to our semi-analytical model for the build-up of the MW, in fact, 
the GM elemental abundances reflect those of MPHS, which are predicted to form out of new virializing haloes 
accreting gas from the GM. This implies that the observed MPHS formed continuously within the redshift range $3 < z \leq 10$. 
From Fig.~\ref{fig:3} we see that our fiducial model yields log~$^6$Li/H$=-13.5$, i.e. about three orders 
of magnitude below the data.
 
This discrepancy cannot be cured by simply boosting the free parameters to their maximum allowed values.
This is also illustrated in the same Figure, where for the upper curve we assume $\epsilon = 1$, 
$\emin = 10$~MeV/n \footnote{This value is exceptionally high and corresponds to the energy at which the \lisix \
production is most efficient. Thus the \lisix \ production will be drastically reduced by increasing $\emin$ 
above this value.} and for the SFR the maximum value allowed by GAMETE within 1-$\sigma$ dispersion.
Although the discrepancy between observations and model results is less prominent in this case, we are 
still unable to fit the data, in particular at [Fe/H]~$=-3$ (i.e. at higher redshifts) only Log~$^6$Li/H$=-12.6$
has had time to be produced, failing short by $30$ times.

In addition the flat data distribution cannot be recovered. It is worth noting that, as also pointed 
out by \cite{Asplund06} \lisix \ may be depleted in stars, mainly during the pre-main sequence phase. If this 
is the case, the \lisix \ abundance observed in stars would not be representative of the gas from which they have 
formed. Taking into account this effect the inferred \lisix \ abundances become metallicity dependent, i.e. the 
flatness is lost. Because of depletion however, the derived \lisix \ values would be higher for all [Fe/H], making 
the discrepancy between our results and observations even larger. 

We finally note, as already claimed by \cite{Rollinde06}, that the production of \liseven \ through this
mechanism is comparable with that of \lisix \ , being the production cross sections of the two isotopes 
very similar. No overproduction of \liseven \ is then expected with respect of the BBN-based value.

\section[]{Discussion}

We have pointed out that both the level and flatness of the \lisix \ distribution 
cannot be explained by CR spallation if these particles have been accelerated by SN shocks 
inside MW building blocks. Although previous claims \citep{Rollinde06} of
a possible solution\footnote{Note that their eq. 18 contains an extra $dz/dt$ term} 
invoking the production of \lisix \ in an early burst of PopIII stars 
have been put forward, such scenario is at odd with both the global properties of the MW 
and its halo MPHS.
     
Our model, which follows in detail the hierarchical build-up of the MW and reproduces 
correctly the MDF of the MPHS, predicts a monotonic increase of \lisix \ abundance with
time, and hence with [Fe/H]. Moreover, our fiducial model falls short of three orders of magnitude
in explaining the data; such discrepancy cannot be cured by allowing the free parameters 
($\emin, \epsilon$) to take their maximum (physically unlikely) values.  
Apparently,  a flat \lisix \ distribution appears inconsistent with any (realistic) model for which
CR acceleration energy is tapped from SNe: if so, \lisix \ is continuously produced 
and destruction mechanisms are too inefficient to prevent its abundance to steadily increase along with [Fe/H].

Clearly, the actual picture could be more complex: for example, if the diffusion 
coefficient in the ISM of the progenitor galaxies is small enough, \lisix \ could
be produced {\it in situ} rather than in the more rarefied GM. This process might increase the 
species abundance, but cannot achieve the required decoupling of \lisix \ evolution from the 
enrichment history. 

Alternatively, shocks associated with structure formation might provide an alternative
\lisix \ production channel \citep{Suzuki02}; although potentially interesting as this
mechanisms decouples metal enrichment (governed by SNe) and CR acceleration (due to structure 
formation shocks), the difficulties that this scenario must face are that (i) at the redshifts ($z=2-3$)
at which shocks are most efficient it must be still [Fe/H]$<-3$, and  (ii) MPHS that formed at 
earlier epochs should have vanishing \lisix \ abundance \citep{Prantzos06}. 

If these issues could represent insurmountable problems, then one has to resort to more 
exotic models involving either suitable modifications of BBN or some yet unknown 
production mechanism unrelated to cosmic SF history.

\section*{Acknowledgments}
We thank M. Kusakabe and E. Rollinde for useful discussions. 
We thank the referee, M.~Asplund, for a careful reading and positive comments.

\bibliography{lithium_main,mn2cls/mn-jour}
\bibliographystyle{mn2cls/mn2e}

\bsp

\label{lastpage}

\end{document}